# Atomically thin sheets of lead-free one-dimensional hybrid perovskites feature tunable white-light emission from self-trapped excitons

Philip Klement[1], Natalie Dehnhardt[2], Chuan-Ding Dong[3], Florian Dobener[1], Samuel Bayliff[4], Julius Winkler[2], Detlev M. Hofmann[1], Peter J. Klar[1], Stefan Schumacher[3,5], Sangam Chatterjee[1], and Johanna Heine[2]*

1 Institute of Experimental Physics I and Center for Materials Research (ZfM), Justus Liebig University Giessen, Heinrich-Buff-Ring 16, D-35392 Giessen, Germany

2 Department of Chemistry and Material Sciences Center, Philipps-Universität Marburg, Hans-Meerwein-Straße, D-35043 Marburg, Germany

3 Department of Physics and Center for Optoelectronics and Photonics Paderborn (CeOPP), Paderborn University, Warburger Strasse 100, D-33098 Paderborn, Germany

4 School of Physics and Astronomy at the University of Minnesota, 116 Church Street S.E., Minneapolis, MN 55455, USA

5 Wyant College of Optical Sciences, The University of Arizona, 1630 E. University Blvd., Tucson, AZ 85721-0094, USA

## Abstract

Low-dimensional organic-inorganic perovskites synergize the virtues of two unique classes of materials featuring intriguing possibilities for next-generation optoelectronics: they offer tailorable building blocks for atomically thin, layered materials while providing the enhanced light harvesting and emitting capabilities of hybrid perovskites. Here, we go beyond the paradigm that atomically thin materials require in-plane covalent bonding and report single layers of the one-dimensional organic-inorganic perovskite $[C_7H_{10}N]_3[BiCl_5]Cl$. Its unique 1D-2D structure enables single layers and the formation of self-trapped excitons which show white light emission. The thickness dependence of the exciton self-trapping causes an extremely strong shift of the emission energy. Thus, such two-dimensional perovskites demonstrate that already 1D covalent interactions suffice to realize atomically thin materials and provide access to unique exciton physics. These findings enable a much more general construction principle for tailoring and identifying two-dimensional materials that are no longer limited to covalently bonded 2D sheets.

## Introduction

Layered organic-inorganic perovskites offer considerable potential for solar cells and other optoelectronic devices as they retain the high-performing properties of three-dimensional halide perovskites with high power conversion efficiency and improved long-term and environmental device stability[1]. Conceptually, these layered materials can be understood as cut-outs across different crystallographic directions of the parent perovskite $AMX_3$ (A = small organic or inorganic cation like $Cs^+$ or $CH_3NH_3^+$; M = divalent metal like $Pb^{2+}$ or $Sn^{2+}$; X = Cl, Br or I). This is achieved by using larger organic cations like primary alkyl- or aryl ammonium ions.[2] The resulting 2D structure can be considered an ideal quantum well of one inorganic layer confined by two organic barriers. The optical properties of such quantum wells can be tuned by varying the thickness of the inorganic layer[3], and the freedom in combining the inorganic and organic components of the material offers a rich chemical, structural, and optical tunability[4]. These materials pave the way to the design of 2D materials as the building blocks can be chosen deliberately. Established 2D materials provide an excellent platform for fundamental research and applications including fundamental physical phenomena such as polaritons[5], superconductivity[6], and charge density waves[7] as well as applications such as water purification[8], light-emitting diodes[9], photovoltaics[10], and sensing[11]. This wide range inherently infers that combinations of layered perovskites and 2D materials will provide advanced multi-functional structures combining many properties.

More recent additions to the family of 2D materials include layered 2D perovskites[12–15]. In principle, any layered material with only weak van der Waals forces acting between its comparatively strongly-bound layers can be exfoliated[16]. The limiting factors and problems in the investigation of free-standing single layers of such layered 2D perovskites have been manifold: (1) Early reports suggested that the quantum well is effectively isolated by the organic layers[17], which means that the properties of the single layer and bulk crystal should be the same. Further, (2) strong in-plane interactions in only two dimensions, typically realized by covalent bonds, were deemed mandatory for the formation of free-standing single layers. Even further, (3) the technical difficulty of separating and analyzing individual layers has led to only few reports[13].

Going beyond the state-of-the-art, we successfully combine the concepts of layered perovskites and atomically thin materials to establish a to date inaccessible class of hybrid materials with unique exciton physics, namely self-trapping in low-dimensional materials. We present $[C_7H_{10}N]_3[BiCl_5]Cl$ (**BBC**), as a principle example of single crystalline organic-inorganic hybrid materials that feature only 1D, wire-like covalent interactions within their layers. This surpasses the conventional paradigm of in-plane covalent bonds being mandatory for atomically thin 2D materials. We find ultrathin crystals of excellent quality with spectrally broadband, white-light emission across the visible spectrum originating from self-trapped excitons. Ultrathin sheets reveal an extremely strong shift of the emission energies due to the thickness dependence of the exciton self-trapping. This suggests that single crystalline compounds featuring lamellar supramolecular motifs can be understood as 2D materials, leading toward a general construction principle for identifying and creating tailored functional 2D materials even when no 2D covalent interactions are present within the individual layers.

**Results**

**Synthesis and structural characterization of BBC.** An illustration of the concept of BBC shows the main interactions in the material (Figure 1a). The strong ionic and supramolecular interactions in the inorganic layer and the weak van-der-Waals interactions between individual layers corroborate the 2D nature of the compound. The reaction of benzylamine with a solution of $Bi_2O_3$ in concentrated hydrochloric acid yields large, colorless plates of BBC (Figure 1b). Most plates are rectangular in shape with truncated ends, reflecting the mechanically brittle structure. The product yield of this straight-forward synthesis was 52 %. We have confirmed the single-crystal structure and high crystallinity of BBC by X-ray diffraction (XRD) measurements (Figure S2). BBC crystallizes in the $P2_1/c$ space group of the monoclinic crystal system, and exhibits a preferred orientation along the (100) direction, which reflects the layered structure and plate habit. The out-of-plane spacing between two layers is 1.7 nm. The lattice parameters are summarized in the Supporting Information (Table S1). The crystal structure shows the hybrid nature of the compound (Figure 1c). Benzylammonium cations are arranged in flat layers, with C-H···π interactions between the benzene rings. This enforces the equally flat arrangement of the $[BiCl_5]^{2-}$ chloro-bismuthate chains, resulting in an overall lamellar structure. Additional chloride ions are found above and below the metalate-ion-plane, completing the compound's charge balance. The crystal structure can be viewed as a bulk arrangement of 1D quantum wires with the inorganic benzylammonium cations surrounding individual chloro-bismuthate chains that are based on corner-sharing octahedral $BiCl_6$ units (Figure 1d). The motif in BBC is reminiscent of 2D lead-halide perovskites such as $(BzA)_2[PbCl_4]$[18] and conceptually derived from the cubic perovskite aristotype (Figure 1e). Within the cubic perovskite, each cation A is replaced by two larger organic cations A' such as benzylammonium (BzA) and an additional halide ion $X^-$ for overall charge balance. This way, <100>-oriented layered perovskites of composition $A'_2MX_4$ are derived. Next, a third of the metal atoms in each of the <100> layers is removed, and the charge on each metal is increased from 2+ to 3+ to maintain neutrality. The result is a vacancy ordered <100> perovskite with *cis*-connected $MX_5$ chains and additional halide ions completing each individual layer. BBC is the first ordered example of a group 15 hybrid organic-inorganic halogenido metalate that is described as a vacancy-ordered 2D perovskite of the <100> family[2,19]. Only one singular, yet intrinsically disordered example of a hybrid iodido bismuthate of this family exists to date: $(H_2AEQT)M_{2/3}I_4$ (M = Bi or Sb; AEQT = 5,5'''-bis-(aminoethyl)-2,2':5',2'':5'',2'''-quaterthiophene)[20]; it features statistically disordered vacancies at all metal positions and the use of a diammonium cation prevents exfoliation in this case.

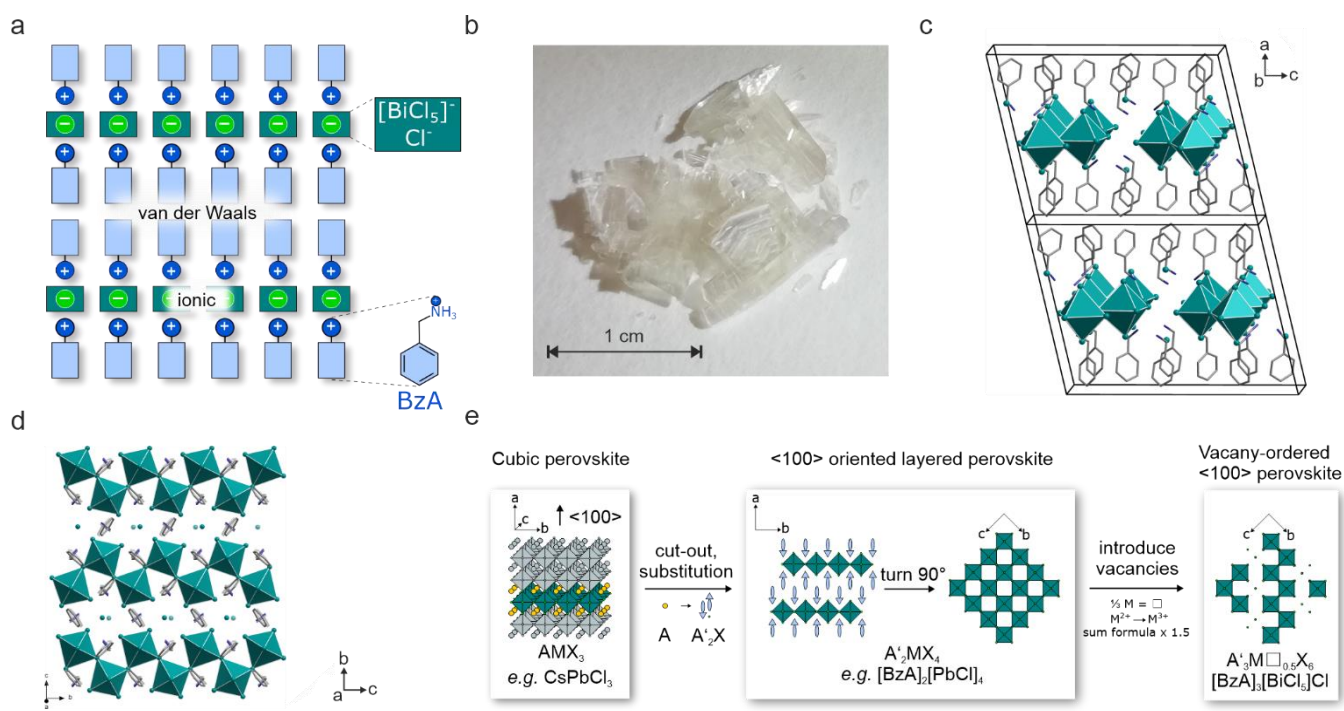

**Figure 1 | Concept, crystal structure and derivation of BBC. a,** Illustration of the layered structure of BBC with strong ionic and supramolecular interactions in the inorganic layer, and weak van der Waals interactions between individual layers. **b,** Photograph of large, colorless BBC crystals. **c,** Layered crystal structure of BBC with Cl atoms in green, C atoms in grey, and H atoms omitted for clarity. **d,** Individual layers are assembled from 1D chloro-bismuthate chains. **e,** Schematic derivation of BBC as a cut-out of the perovskite aristotype.

**Electronic band structure and exfoliation of BBC.** The conventional belief that 2D covalent interactions are an essential prerequisite for stable 2D materials suggests that the mechanical exfoliation should not be possible for BBC[12]. We investigate the mechanical exfoliation process of BBC despite this paradigm. Microscopic modelling evaluates its ability for exfoliation, any possible cleavage plane(s), and compares BBC to well-established 2D materials (Figure S4). The required energy for expansions of the unit cell along the three crystallographic axes indicates the solute binding of BBC along the *a* axis and determines the cleavage plane to the *b*-*c*-plane of the crystal. The calculated crystal structure of the single layer shows only a slight rearrangement of the organic molecules as the unit cell is expanded along the **a** axis. This predicts excellent preservation of the individual sheets upon exfoliation.

The density functional theory (DFT) calculations also allow insight into the electronic properties of BBC. The calculated density of states exhibits a band gap energy of 3.62 eV (Figure 2a), which increases to 3.67 eV when the thickness is reduced to a single layer (Figure 2b). The corresponding band structures show very narrow bandwidths and direct gaps located at the Γ point (Figure S5a-b). The flat bands indicate that the electronic states are highly localized. Only the conduction band shows some dispersion around the Γ point indicating delocalization along the wires. The charge density map of the conduction band (CB) confirms that the states are confined to the $[BiCl_5]^{2-}$ anions (Figure 2c), and mainly contributed by the Bi and Cl orbitals in the calculated density of states (Figure S5d-e). The localization of the valence band slightly increases from the bulk crystal to the single layer. In the bulk crystal, the bands are formed by hybridization of C orbitals from the benzylammonium cations with Cl orbitals from the $[BiCl_5]^{2-}$ anions as suggested by the charge density map of the valence band (VB) (Figure S5d). In the single layer, the C orbitals shift to lower energies leaving the exclusive contribution by Cl orbitals and consequent localization to the $[BiCl_5]^{2-}$ anions. We find no indication for lattice distortion. These results indicate strong charge localization in the inorganic chains of BBC with neglectable interchain coupling, especially in the single layer. Therefore, upon photoexcitation each chloro-bismuthate chain can act as an efficient 1D luminescence center by trapping of excitons.

We have verified the existence of free-standing single layers of BBC by mechanically exfoliating single crystals yielding ultrathin sheets down to the single layer. The procedure is similar to the technique that is successfully applied to common van-der-Waals materials. Typical ultrathin BBC sheets on $SiO_2$/Si substrates exhibit lateral dimensions of a few microns - similar to other 2D materials (Figure 2d). Various samples of only a few unit cells thickness have been realized this way featuring reduced lateral dimensions (less than 1 µm). The structure appears mechanically brittle. The comparison of the expansion energies of the 2D sheets of BBC and common 2D materials like graphene and molybdenum disulfide ($MoS_2$) reveals that significantly less energy is required to expand the 2D sheet of BBC explaining our observations (Figure S4). The ultrathin sheets are reasonably stable in air - at least on the order of days - and long-term stable in inert atmospheres like Ar and in vacuum, where they sustain repeated thermal cycling to cryogenic temperatures without damage. Degradation after 15 weeks of storage under ambient conditions becomes apparent by a reduced thickness and a deterioration of the surface topography (Figure S6). Optical bright-field microscopy and atomic-force microscopy of ultrathin sheets reveal smooth and uniform surfaces highlighting the good crystal quality (Figure 2e). The thicknesses of the sheets range from several hundreds to only a few nanometers. The thinnest sheets are 3 nm high (Figure 2f) which corresponds to a single layer of BBC as this is significantly less than twice the corresponding layer spacing along the *a*-axis of the bulk crystal. Apparently, substrate interactions increase the observed thickness of the first attached layer[12,21].

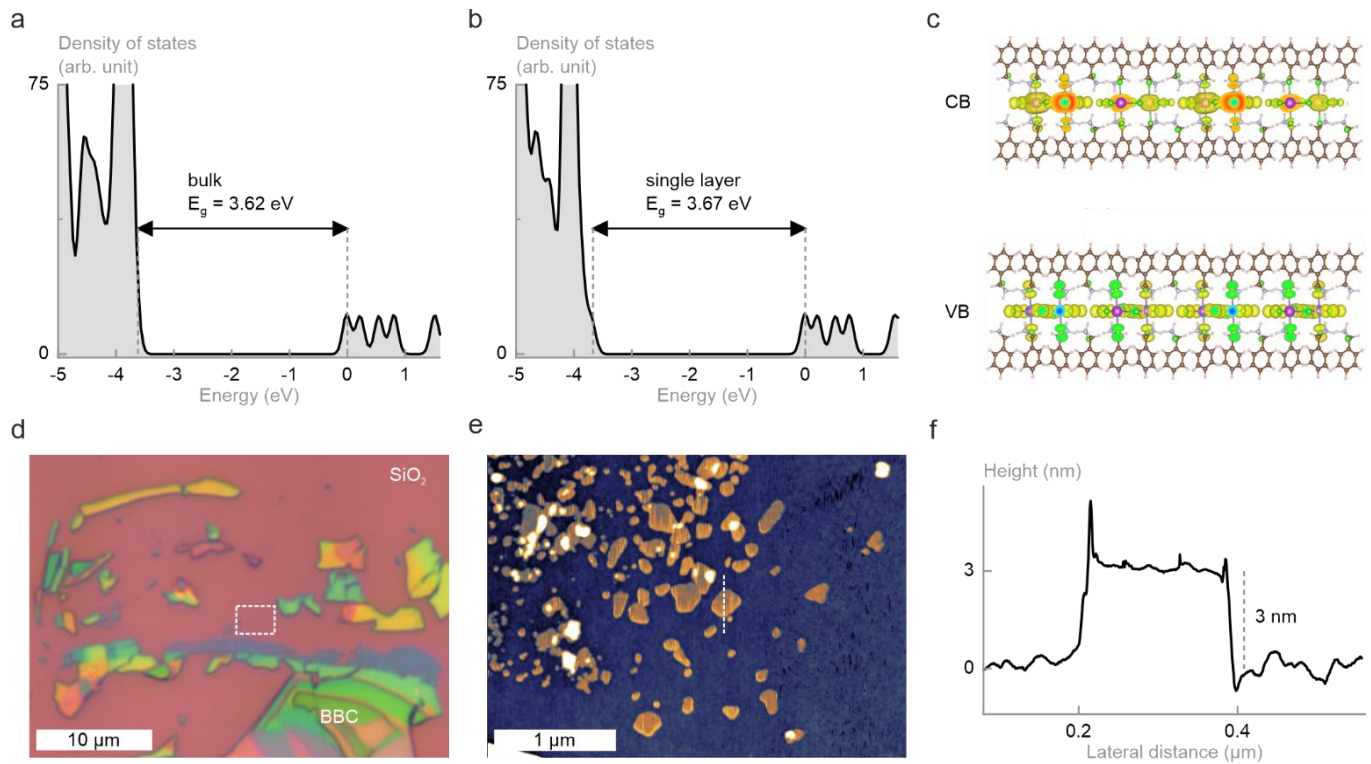

**Figure 2 | Electronic band structure and exfoliation of BBC. a-b**, Computed density of states of bulk and single layer BBC. Bulk and single layer BBC exhibit band gaps of 3.62 and 3.67 eV. The conduction band peaks are aligned to 0 energy for better comparability. **c**, Conduction band (CB) and valence band (VB) associated with charge density maps of single layer BBC. CB and VB are mainly confined to the inorganic chloro-bismuthate chains. **d**, Typical bright-field microscopy image of mechanically exfoliated ultrathin BBC sheets on a SiO₂/Si substrate. **e**, AFM image of the area highlighted by the white rectangle in **c** showing single layers (orange). **f**, Height profile of a single layer highlighted by the white dashed line in **d**. The thickness of the single layer is 3 nm.

**Thickness-dependent optical properties of ultrathin BBC.** BBC displays strong, broadband emission across the visible spectrum originating from self-trapped excitons (STEs). The absorption spectrum (Figure 3a) shows the onset of absorption at 3.3 eV (375 nm), and a resonance at 3.5 eV (354 nm), which we identify with the fundamental bright exciton resonance. The increased band gap energy compared to 3D perovskites is assigned to the increased quantum confinement in the 1D structure. The photoluminescence (PL) of bulk BBC crystals centers at $E_{STE}$ = 1.87 eV (663 nm) with a large full width at half maximum (FWHM) of 0.53 eV (185 nm) making BBC a potential broadband emission material. The large Stokes shift of 1.6 eV (466 nm) is among the highest values reported for any solid-state light-emitting material[22]. Broadband emissions with large Strokes shifts are typically associated with STE, that form in materials with deformable lattice and localized carriers[23–25], *i.e.,* strong electron-phonon coupling and the absence of deep defects within the band gap[26]. An additional, much weaker PL feature with a comparatively small FWHM appears near 3 eV. We attribute this to the free exciton transition (FX), which is Stokes-shifted by 0.5 eV due to the strong electron-phonon coupling in the material.

We quantify the electron-phonon coupling by deriving the Huang-Rhys factor *S* from the temperature dependence of the PL within a configuration coordinate model (CCM)[24]. While the peak PL emission energy shifts slightly towards lower energies with increasing lattice temperature, its bandwidth extends considerably with the FWHM increasing from 0.53 to 0.73 eV (185 to 260 nm) yielding a large *S* = 45 with the phonon frequency of the excited state

$\hbar\Omega_e$ = 32 meV (Figure S7a-b). From these data, we estimate the exciton self-trapping time to $\tau = 2\pi/\Omega_e$ = 129 fs, which indicates the ultrafast STE formation. The temperature dependence of the absorption yields the phonon frequency of the ground state $\hbar\Omega_g$ = 19 meV (Figure S7c), and the lattice deformation energy $E_d = S\hbar\Omega_g$ = 0.86 eV. We confirm the absence of deep defects within the band gap by photoluminescence excitation (PLE) spectroscopy of the white light emission at 1.9 eV (Figure 3b). The PLE intensity reveals a strong exciton peak at 3.4 eV followed by a continuum contribution peaking at 3.7 eV. The shape of the exciton peak is similar to low-dimensional Wannier-like absorption patterns[27,28] and reflects the low-dimensional structure of BBC[4]. The energy difference of the exciton resonance and the continuum of 0.3 eV yields an estimation of the exciton binding-energy of 10 to 75 meV for mixed dimensional confinement[4]. Importantly, the PLE intensity decreases to nearly zero below the band gap, and we observe no PLE intensity near 1.9 eV. A comparison of PLE spectra of the STE and FX emission reveal identical shapes and features indicating that the white light emission originates from the same excited state (Figure S7e). Further, the PL intensity increases linearly with the excitation power in the investigated range (Figure S7f), indicating that there is neither a threshold for the onset of PL due to non-radiative recombination through defects at low excitation powers nor saturation of the absorption at high excitation powers or multi-photon absorption. These experimental observations of strong electron-phonon coupling and absence of any indications of strong defect emission confirm the origin of the white light emission as STE rendering BBC a 2D material candidate with functionality as white-light source. We measured the PL lifetime of the STE state by time-resolved PL (Figure 3c). The transient shows a bi-exponential decay after the laser excitation with lifetimes of $\tau_1 = 33$ ps and $\tau_2 = 710$ ps, which are comparable to other few-layered Pb-based 2D perovskites[12,15].

These experimental findings allow us to establish an energy-level diagram of BBC based on the configuration coordinate model (Figure 3d). After photoexcitation at or above the exciton absorption resonance energy $E_g$ = 3.4 eV, we find emission at $E_{FX}$ = 3 eV and an ultrafast transfer to the self-trapped state which is calculated to 129 fs. The observation of a self-trapped state is consistent with the deformable lattice of BBC and the absence of a potential barrier for the formation of STE in 1D structures as opposed to 3D systems[23]. The self-trapping energies and the lattice deformation energies are the differences of the excited-state energies and ground-state energy between the STE and FX and amount to $E_{st}$ = 0.27 eV and $E_d$ = 0.86 eV, respectively. STEs recombine after the PL lifetime of 710 ps with the emission energy $E_{STE} = E_{FX} - E_b - E_{st} - E_d$ = 1.87 eV. The zero-phonon line (ZPL) $E_{ZPL} = E_{STE} + S\hbar\Omega_e$ = 3.1 eV from the CCM coincides with the narrow PL feature at 3 eV confirming the identification with the free exciton. Importantly, the 1D structure of BBC allows for STE formation and the confinement of electrons and holes to the inorganic chain renders the STE transition radiative. Thus, BBC can be considered as a potential white light emitter.

We examine the evolution of the characteristic lattice vibrations of the bulk crystal and several ultrathin sheets by resonant and non-resonant Raman spectroscopy (Figure S8). The non-resonant Raman spectra are typical for 1D structures with an octahedral arrangement of halogen atoms with two chlorine bridges linking two adjacent bismuth atoms. We attribute the peaks at 284, 249, 237, and 161 cm$^{-1}$ to Bi-Cl stretching vibrations (2A$_1$, B$_1$, E$_1$) and the peaks at 106 and 80 cm$^{-1}$ to Ci-Bi-Cl bending vibrations[29]. The intensity of the peaks decreases with the sample thickness and eventually vanishes for ultrathin sheets indicating weak Raman scattering efficiencies. We find that the Raman spectra change significantly between resonant and non-resonant excitation. This provides corroborating yet more direct experimental evidence for selftrapping as we find that optical excitation deforms the lattice.

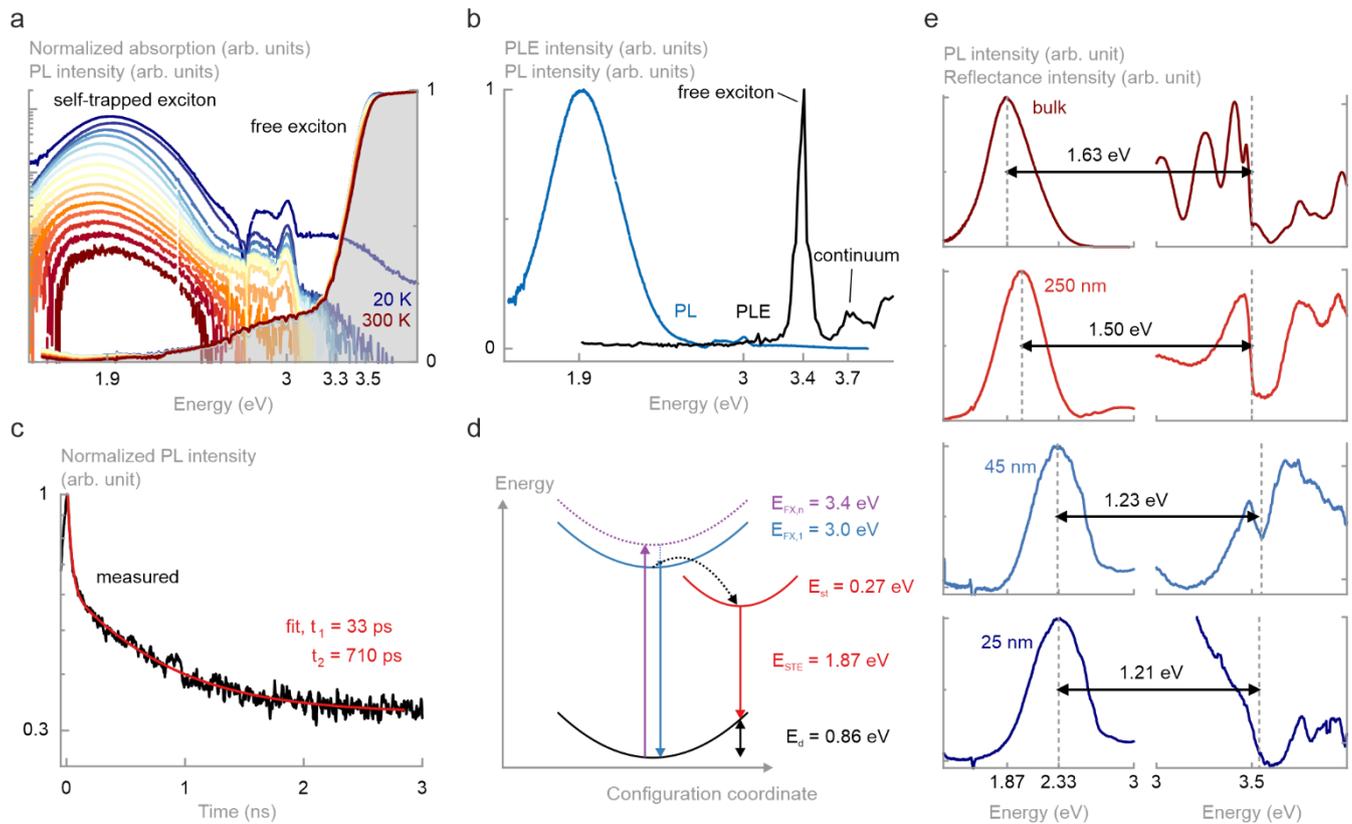

**Figure 3 | Thickness-dependent optical properties of self-trapped excitons in BBC. a**, Absorption and photoluminescence (PL) of a bulk crystal as a function of temperature. PL spectra exhibit white light emission across the visible spectrum from self-trapped excitons (STE) and a narrow emission from free excitons (FX). **b**, Photoluminescence excitation and PL spectra of a bulk crystal showing no defect-related emission. **c**, Time-resolved PL of the STE emission near 1.9 eV of a bulk crystal showing a luminescence lifetime of 710 ps. **d**, Configuration coordinate diagram of the STE and FX formation in BBC. The arrows represent lattice relaxations and optical transitions. **e**, Thickness-dependent PL and differential reflectance spectra of ultrathin BBC. The STE emission shifts to higher energies by 0.46 eV when the thickness decreases, whereas the band gap only shifts by 0.04 eV.

The unique combination of the chain-like 1D structure of BBC with the formation of STE and the layered crystal structure with exfoliation of ultrathin sheets make the investigation of the thickness dependence of the STE possible. We measure the PL and differential reflectance spectra of ultrathin sheets of BBC as a function of the sample thickness (Figure 3e). Bulk crystals and ultrathin sheets display spectrally broadband emission. The emission energy shifts by as much as 0.46 eV to higher energies from 1.87 (663 nm) to 2.33 eV (532 nm) as the sample thickness reduces, while the FWHM stays constant. The occurrence and the magnitude of the shift are highly remarkable as the latter is at least one order of magnitude larger than that found for FX-related PL in 2D hybrid perovskites[12,13,30]. This presumably represents the first observation of a thickness-dependent shift of the STE emission in a perovskite material. We identify its origin from the variation of the differential reflectance: the band gap energy shifts only slightly to higher energies by 0.04 eV from 3.50 (354 nm) to 3.54 eV (350 nm) in stark contrast to the STE PL energy. This is consistent with the results from our DFT calculations that show only a slightly increased carrier localization in the single layer (Figure 2 & S5) compared to the bulk. The Stokes shift, which is the sum of the exciton binding energy, self-trapping energy, and lattice deformation energy, reduces from 1.63 eV in the bulk crystal to 1.21 eV in the ultrathin sheet. We exclude increased quantum confinement and reduced dielectric screening in the ultrathin sheets as the origin of the STE PL shift as the band gap is almost independent of the sample thickness in our experiments, and the exciton binding energy was found to be independent of the sample thickness as well[13,15]. Consequently, the large STE PL shift is directly related to the thickness dependence of the STE and its vibrational coupling. Both, the self-trapping energy and the lattice deformation energy decrease in ultrathin samples shifting the STE emission to higher energies rendering self-trapping of excitons weaker in the single layer limit. The 1D hybrid material BBC thus provides an excellent platform for fundamental research and promising applications. The thickness dependence of the exciton self-trapping down to the single-layer limit is of fundamental interest. It may be used to tune the emission color by simply changing the crystal or film thickness. We anticipate that these results will encourage research on novel 1D hybrid materials for next generation lightning technologies.

**Conclusions**

In conclusion, we successfully combine the concepts of layered perovskites and atomically thin materials and establish a new class of hybrid materials with unique exciton physics. The existence of single layers of [$C_7H_{10}N$]$_3$[$BiCl_5$]Cl (**BBC**) contradicts the apparent paradigm that atomically thin 2D materials require only 2D covalent interactions. BBC is a vacancy-ordered, layered perovskite of the <100> family and features only 1D, wire-like covalent interactions within its layers. The ionic and supramolecular interactions present in this compound are strong enough to allow for mechanical exfoliation of bulk crystals down to single layers. We find ultrathin crystals of excellent quality with white light emission across the visible spectrum originating from self-trapped excitons (STE) that form due to the 1D structure. Ultrathin sheets reveal an extremely strong shift of the emission due to the thickness dependence of the exciton self-trapping, which allows the facile color tuning of white light emitters by changing the film thickness. These findings enable a general construction principle for identifying and creating tailored, custom-functionalized 2D materials where specific functionalities are introduced both in the organic and the inorganic parts of the materials. This class of materials enables interface-controlled device integration of brightly luminescent 1D and 0D hybrid perovskites and offers a promising pathway for the non-covalent functionalization of classical 2D materials through heterostructures.

**Methods**

**Synthesis.** $Bi_2O_3$ (233 mg, 0.5 mmol) was dissolved in 10 mL of concentrated hydrochloric acid (37 %). A stoichiometric amount of benzylamine (0.33 mL, 3 mmol) was added. The solution was stirred and heated to reflux for 30 min, then left to cool to room temperature undisturbed. [BzA]$_3$[BiCl$_5$]Cl (193 mg, 0.26 mmol, 52 %) was obtained as large colourless crystals. CHN Data: Anal. Calcd for $C_{21}H_{30}BiCl_6N_3$, (M = 746.16 g mol$^{-1}$): C, 33.80; H, 4.05; N, 5.63 %. Found: C, 33.83; H, 4.06; N, 5.64 %.

**Single crystal X-ray data brief.** Monoclinic, space group $P2_1/c$ (no. 14), $a$ = 17.0182(8) Å, $b$ = 7.6205(4) Å, $c$ = 22.4391(9) Å, $\beta$ = 102.638(2)°, $V$ = 2839.6(2) Å$^3$, $Z$ = 4, $T$ = 100 K, $R_1$ = 0.0565 (I > 2σ(I)), $wR_2$ = 0.1268 (all data).

Additional details on data collection and structure refinement can be found in the Supplementary Information. Structure data has also been deposited in the Cambridge Crystallographic Data Centre under accession code CCDC 1952283.

**Modelling.** Density functional theory (DFT) calculations were performed using projector-augmented wave (PAW) potentials and plane-wave basis sets, as implemented in the VASP software package. The PBE functional and Tkatchenko-Scheffler method were used to describe exchange-correlation and dispersion interaction, respectively. HSE06 hybrid functional was used to calculate the density of states as well as the partial charge densities with spin-orbital coupling (SOC) included. In addition, band structure calculations were performed by using PBE functional with SOC.

**Mechanical exfoliation and AFM.** Ultrathin sheets of [BzA]$_3$[BiCl$_5$]Cl were produced by exfoliation of bulk single crystals onto Si substrates with a 275 nm thick wet thermal oxide layer. Thin sheets were identified by optical contrast imaging and confirmed by Atomic Force Microscopy (AFM, AIST-NT SmartSPM 1000). Noncontact tapping mode was used to avoid sample damage.

**Optical measurements.** All optical measurements were conducted at cryogenic Helium temperatures with the samples in vacuum. For PL and µ-PL measurements, samples were excited at 266 nm (4.66 eV). The spot diameters and excitation powers were 200 µm, 65 µW, and 1 µm, 1.4 µW, respectively. Radiation from a combined deuterium and tungsten lamp source was used for absorption and µ-reflectance measurements (spot diameters of 300 and 1 µm, respectively). For PLE measurements, samples were excited with 50 fs pulses at a fluence of (2200 ± 375) µJ cm$^{-2}$. For µ-TRPL measurements, a samples were excited with 120 fs pulses at 266 nm (1 µm spot diameter, 1.4 µW).

Detailed methods are given in the Supplementary Information.

**Acknowledgements.**

This work is funded by the German Research Foundation (DFG) via the collaborative research center SFB 1083. S. C. acknowledges financial support by the Heisenberg programme (CH660/8). C.-D.D. and S.S. acknowledge funding through project SCHU 1980/13 and through the Heisenberg programme (No. 270619725). A grant for computing time at the Paderborn Center for Parallel Computing (PC$^2$) is also gratefully acknowledged. N. D. thanks the Fonds der Chemischen Industrie and the Studienstiftung des Deutschen Volkes for their support. S. B. thanks the DAAD / RISE program for a research internship. J. H. thanks Prof. Stefanie Dehnen for her constant support.

**Author contributions.**

J.H. and S.C. conceived the idea, and P.K assisted in the design of the experiments. N.D., S.B., and J.W. developed and optimized the synthesis and performed structural characterizations. J.H. performed the XRD characterizations. C.-D.D. and S.S. performed the theoretical modelling. P.K. prepared the ultrathin samples and performed the AFM, absorption, reflectance, PL, TRPL, and Raman characterizations; F.D. performed the PLE characterization. P.K. and S.C. analyzed the optical data with input from F.D., D.M.H., and P.J.K.. P.K, C.-D.D., S.S., S.C. and J.H. wrote the paper. All authors commented on the manuscript.

**Competing interests.**

The authors declare no competing interests.

**References**

1. Etgar, L. The merit of perovskite's dimensionality; Can this replace the 3D halide perovskite? *Energy Environ. Sci.* **11**, 234–242 (2018).

2. Saparov, B. & Mitzi, D. B. Organic-Inorganic Perovskites: Structural Versatility for Functional Materials Design. *Chemical Reviews* **116**, 4558–4596 (2016).

3. Blancon, J.-C. et al. Extremely efficient internal exciton dissociation through edge states in layered 2D perovskites. *Science* **355**, 1288–1292 (2017).

4. Straus, D. B. & Kagan, C. R. Electrons, Excitons, and Phonons in Two-Dimensional Hybrid Perovskites: Connecting Structural, Optical, and Electronic Properties. *J. Phys. Chem. Lett.* **9**, 1434–1447 (2018).

5. Low, T. et al. Polaritons in layered two-dimensional materials. *Nat. Mater.* **16**, 182–194 (2017).

6. Fatemi, V. et al. Electrically tunable low-density superconductivity in a monolayer topological insulator. *Science* **362**, 926–929 (2018).

7. Xi, X. et al. Strongly enhanced charge-density-wave order in monolayer $NbSe_2$. *Nat. Nanotechnol.* **10**, 765–769 (2015).

8. Teng, Z. et al. Edge-Functionalized $g-C_3N_4$ Nanosheets as a Highly Efficient Metal-free Photocatalyst for Safe Drinking Water. *Chem* **5**, 664–680 (2019).

9. Ross, J. S. et al. Electrically tunable excitonic light-emitting diodes based on monolayer $WSe_2$ p-n junctions. *Nat. Nanotechnol.* **9**, 268–272 (2014).

10. Pospischil, A., Furchi, M. M. & Mueller, T. Solar-energy conversion and light emission in an atomic monolayer p-n diode. *Nat. Nanotechnol.* **9**, 257–261 (2014).

11. Klement, P., Steinke, C., Chatterjee, S., Wehling, T. O. & Eickhoff, M. Effects of the Fermi level energy on the adsorption of $O_2$ to monolayer $MoS_2$. *2D Mater.* **5**, (2018).

12. Dou, L. et al. Atomically thin two-dimensional Organic-inorganic hybrid perovskites. *Science* **349**, 1518–1521 (2015).

13. Yaffe, O. et al. Excitons in ultrathin organic-inorganic perovskite crystals. *Phys. Rev. B - Condens. Matter Mater. Phys.* **92**, 1–7 (2015).

14. Yang, S. et al. Ultrathin Two-Dimensional Organic–Inorganic Hybrid Perovskite Nanosheets with Bright, Tunable Photoluminescence and High Stability. *Angew. Chemie - Int. Ed.* **56**, 4252–4255 (2017).

15. Zhang, Q., Chu, L., Zhou, F., Ji, W. & Eda, G. Excitonic Properties of Chemically Synthesized 2D Organic–Inorganic Hybrid Perovskite Nanosheets. *Adv. Mater.* **30**, 1–8 (2018).

16. Nicolosi, V., Chhowalla, M., Kanatzidis, M. G., Strano, M. S. & Coleman, J. N. Liquid exfoliation of layered materials. *Science* **340**, 72–75 (2013).

17. Ishihara, T., Takahashi, J. & Goto, T. Optical properties due to electronic transitions in two-dimensional semiconductors $(C_nH_{2n+1}NH_3)_2PbI_4$. *Phys. Rev. B* **42**, 11099–11107 (1990).

18. Liao, W. Q. et al. A lead-halide perovskite molecular ferroelectric semiconductor. *Nat. Commun.* **6**, 1–7 (2015).

19. Mao, L., Stoumpos, C. C. & Kanatzidis, M. G. Two-Dimensional Hybrid Halide Perovskites: Principles and Promises. **3**, (2019).

20. Mitzi, D. B. Organic-inorganic perovskites containing trivalent metal halide layers: The templating influence of

the organic cation layer. *Inorg. Chem.* **39**, 6107–6113 (2000).

21. Shearer, C. J., Slattery, A. D., Stapleton, A. J., Shapter, J. G. & Gibson, C. T. Accurate thickness measurement of graphene. *Nanotechnology* **27**, (2016).

22. Zhou, C. *et al.* A Zero-Dimensional Organic Seesaw-Shaped Tin Bromide with Highly Efficient Strongly Stokes-Shifted Deep-Red Emission. *Angew. Chemie - Int. Ed.* **57**, 1021–1024 (2018).

23. Song, K. S. & Williams, R. T. *Self-Trapped Excitons*. (Springer Berlin Heidelberg, 1993).

24. Stadler, W. *et al.* Optical investigations of defects in $Cd_{1-x}Zn_xTe$. *Phys. Rev. B* **51**, 10619–10630 (1995).

25. Luo, J. *et al.* Efficient and stable emission of warm-white light from lead-free halide double perovskites. *Nature* **563**, 541–545 (2018).

26. Kahmann, S., Tekelenburg, E. K., Duim, H., Kamminga, M. E. & Loi, M. A. Extrinsic nature of the broad photoluminescence in lead iodide-based Ruddlesden–Popper perovskites. *Nat. Commun.* **11**, 1–8 (2020).

27. Elliott, R. J. Intensity of optical absorption by excitons. *Phys. Rev.* **108**, 1384–1389 (1957).

28. Tanguy, C., Lefebvre, P., Mathieu, H. & Elliott, R. J. Analytical model for the refractive index in quantum wells derived from the complex dielectric constant of Wannier excitons in noninteger dimensions. *J. Appl. Phys.* **82**, 798–802 (1997).

29. Cariati, F. *et al.* Solid state vibrational spectroscopy of some piperidinium and morpholinium salts of tin(IV), antimony(III) and bismuth(III) halide complexes. *Spectrochim. Acta Part A Mol. Spectrosc.* **34**, 801–805 (1978).

30. Zhang, R. *et al.* Air-Stable, Lead-Free Zero-Dimensional Mixed Bismuth-Antimony Perovskite Single Crystals with Ultra-broadband Emission. *Angew. Chemie - Int. Ed.* **58**, 2725–2729 (2019).

**Extended Methods**

**Modelling.** Density functional theory (DFT) calculations were performed using projector-augmented wave (PAW) potentials[1] and plane-wave basis sets, as implemented in the VASP software package.[2,3] The PBE functional[4] and a plane-wave basis set with energy cut off 550 eV were used to optimize the geometries. Tkatchenko-Scheffler method[5] is adopted for the dispersion interaction. The crystallographic unit cell (Table S1) was used as the starting point for the lattice expansion calculations, where the lattice constants ($a$, $b$, and $c$) were expanded individually up to 20 % in steps of 2 %, and at each point the geometry was optimized for fixed lattice constants. For the electronic structure calculations, the crystallographic unit cell (Table S1) and a 2*4*2 k-point lattice were used for bulk BBC, while for single layer BBC the lattice constant $a$ was increased to 30 Å to introduce sufficient interlayer vacuum space, and a 1*4*4 k-point lattice was used. HSE06 hybrid functional was used to calculate the density of states as well as the partial charge densities with spin-orbital coupling (SOC) included. In addition, band structure calculations were performed by using PBE functional with SOC.

**Optical measurements.** Absorption and PL measurements were conducted in a closed-cycle He cryostat from 20 to 300 K with the sample in vacuum. Radiation from a combined deuterium and tungsten lamp source was used for absorption measurements. For PL measurements, excitation light from an Ar-ion laser operating at a wavelength of 266 nm (4.66 eV) was focused into a 200 µm diameter spot with 65 µW excitation power. The absorption and PL signals were spectrally filtered through a 0.32 m monochromator (Horiba Triax 320) and detected on a Si charge-coupled device camera (Andor DU420). PLE, µ-TRPL, µ-PL and µ-reflectance measurements were conducted in a continuous-flow liquid He cold finger cryostat at cryogenic Helium temperature with the sample in vacuum. For PLE measurements, a Ti:sapphire regenerative amplifier at a 5 kHz repetition rate (Spectra Physics Solstice ACE) and a pulse length of 50 fs in combination with an optical parametric amplifier (Lightconversion Topas + UV-VIS extension) was used. The fluence was (2200 ± 375) µJ cm$^{-2}$. Dielectric short-pass in the excitation beam allowed suppressing non-linear mixing residues from the laser system. The PLE signal was spectrally filtered through a 193 mm monochromator (Andor Kymera 193i) and detected by a Si charge-coupled device camera (Andor DU401). Spectral integration from 600 to 1086 nm (1.14 to 2.1 eV) reduced the 2D PLE signal to a one-dimensional PLE spectrum. For µ-TRPL measurements, a Ti:sapphire laser emitting 120 fs pulses at a repetition rate of 78 MHz at 800 nm (1.55 eV) was frequency-doubled (400 nm, 3.1 eV) and frequency tripled for excitation (266 nm, 4.66 eV). For µ-PL measurements, excitation light from an Ar-ion laser operating at a wavelength of 266 nm (4.66 eV) was used. Radiation from a deuterium lamp source was used for µ-reflectance measurements. The excitation light was focused into a 1 µm diameter spot using a confocal beam path with all-reflective optics. The excitation power was 1.4 µW in µ-TRPL and 20 µW in µ-PL measurements. The PL signal was spectrally filtered through a 1/4 m monochromator (Oriel MS260i) and detected by a Si charge-coupled device camera (Andor iDUS 420). The time-resolved data was recorded by a streak camera equipped with a S20 cathode yielding time resolution better than 1 ps (Hamamatsu C10910 and Orca Flash).

**Crystallographic Details**

**Table S1:** Crystallographic data of [BzA]$_3$[BiCl$_5$]Cl (**BBC**, BzA = benzylammonium, C$_7$H$_{10}$N$^+$)

| | BBC (CCDC 1952283) |
|---|---|
| Empirical formula | C$_{21}$H$_{30}$BiCl$_6$N$_3$ |
| Formula weight /g·mol$^{-1}$ | 746.16 |
| Crystal color and shape | Colorless plate |
| Crystal size | 0.315 × 0.271 × 0.092 |
| Crystal system | monoclinic |
| Space group | $P2_1/c$ |
| $a$ /Å | 17.0182(8) |
| $b$ /Å | 7.6205(4) |
| $c$ /Å | 22.4391(9) |
| $\alpha$ /° | 90 |
| $\beta$ /° | 102.638(2) |
| $\gamma$ /° | 90 |
| $V$ /Å$^3$ | 2839.6(2) |
| $Z$ | 4 |
| $\rho_{calc}$ /g·cm$^{-3}$ | 1.745 |
| $\mu$(Mo$_{K\alpha}$) /mm$^{-1}$ | 6.788 |
| measurement temp. /K | 100 |
| Absorption correction type | multi-scan |
| Min/max transmission | 0.384/0.745 |
| $2\Theta$ range /° | 4.884-50.562 |
| No. of measured reflections | 24482 |
| No. of independent reflections | 5098 |
| $R$(int) | 0.0917 |
| No. of indep. reflections ($I > 2\sigma(I)$) | 4354 |
| No. of parameters | 246 |
| $R_1$ ($I > 2\sigma(I)$) | 0.0565 |
| $wR_2$ (all data) | 0.1268 |
| $S$ (all data) | 1.065 |
| $\Delta\rho_{max}$, $\Delta\rho_{min}$ /e· Å$^{-3}$ | 2.59/-2.67 |

Single crystal X-ray determination was performed on a Bruker Quest D8 diffractometer with microfocus MoKα radiation and a Photon 100 (CMOS) detector. Data collection and processing including the twin integration was performed with the Bruker APEX software package.[6] The structure was solved using direct methods, refined by full-matrix least-squares techniques and expanded using Fourier techniques, using the Shelx software package[7] within the OLEX2 suite.[8] All non-disordered, non-hydrogen atoms were refined anisotropically. Hydrogen atoms were assigned to idealized geometric positions and included in structure factors calculations. Pictures of the crystal structure were created using DIAMOND.[9]

Due to the layered nature of the compound twinning was observed and accounted for by integration of two separate domains. This also necessitated the use of ISOR, SADI and RIGU restraints on the organic part of the crystal structure during refinement to obtain reasonable displacement parameters. The free chloride ion in the crystal structure is found to be disordered over two positions. Occupancies were first refined freely, then constrained to the appropriate values (0.55 and 0.45) to stabilize the refinement. The disordered atoms were refined isotropically.

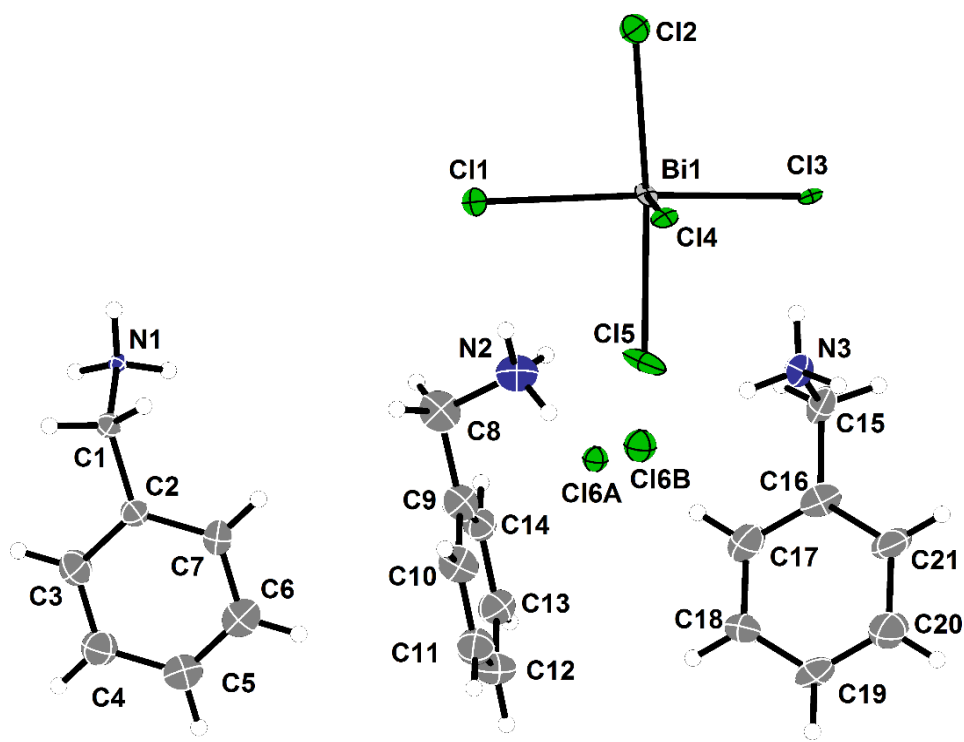

**Figure S1 | Asymmetric unit.** Asymmetric unit of the single crystal structure of BBC, ellipsoids shown at 70%. Bi-Cl bond length range from 2.587(3) to 2.869(2) Å, typical for chlorido bismuthates.[10]

**Powder diffraction**

Powder patterns were recorded on a STADI MP (STOE Darmstadt) powder diffractometer, with CuKα1 radiation with λ = 1.54056 Å at room temperature in transmission mode. The patterns confirm the presence of the phase determined by SCXRD measurements and the absence of any major crystalline by-products. Patterns were simulated from single crystal data using Mercury.[11]

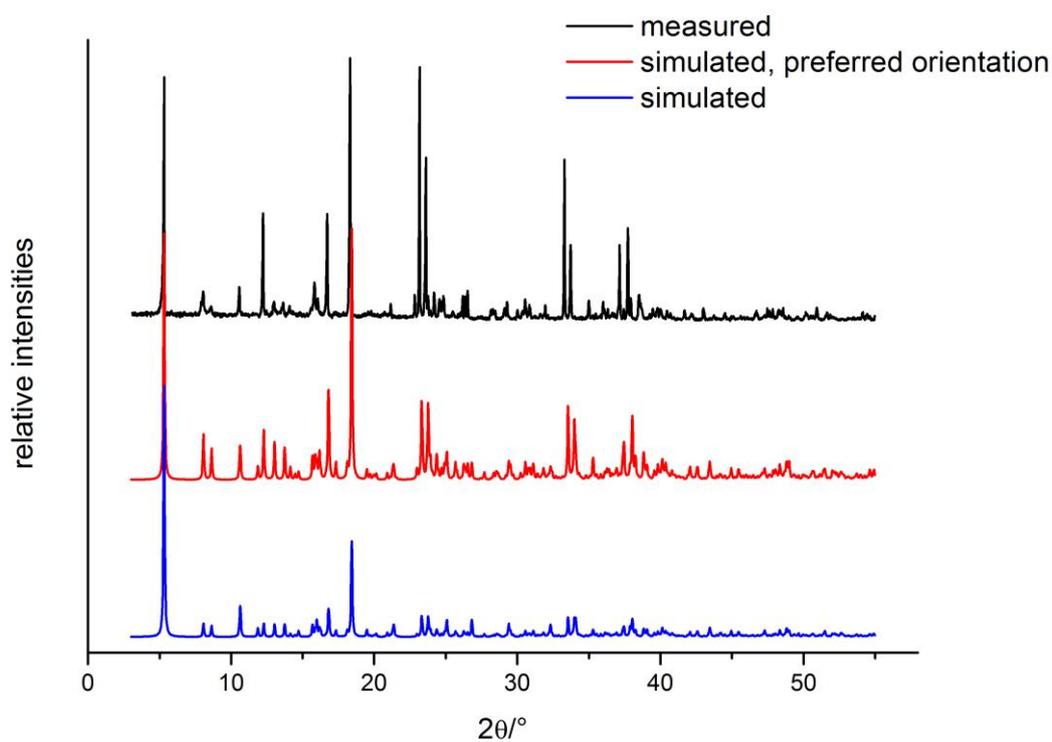

**Figure S2 | XRD.** Measured and simulated powder patterns of BBC. The diffractogram shows signs of texture effects, which can be demonstrated by simulating a diffractogram with a (100) preferred orientation, as can be expected from the layered crystal structure and plate-like crystal habit.

## Thermal analysis

The thermal behavior of BBC (16.8 mg), was studied by simultaneous DTA/TG on a NETZSCH STA 409 C/CD from 25 °C to 800 °C with a heating rate of 10 °C min$^{-1}$ in a constant flow of 80 ml min$^{-1}$ Ar.

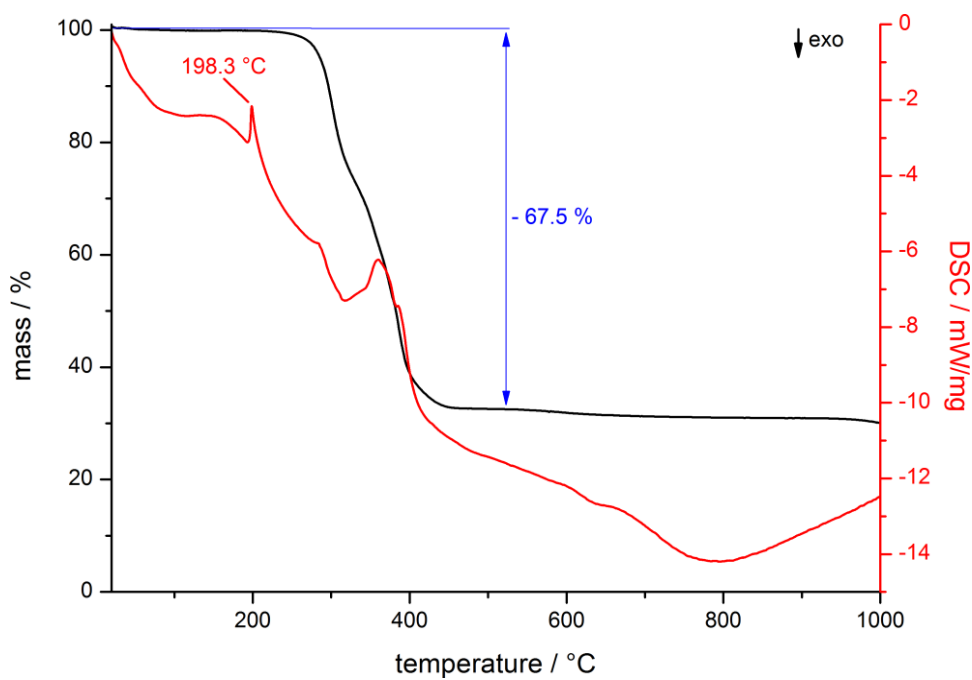

**Figure S3 | Thermal analysis.** A melting point or solid-solid phase transition can be observed at 198.3 °C and the onset of decomposition (1 % mass loss) at 257.1 °C.

## Bi-Cl bond length upon expansion of the lattice constants

Examination of the change in the Bi-Cl bond length of the inorganic Bi-Cl-octahedral reveals similar results for expansions along the **b** axis and the **c** axis (Figure S4a). Consequently, the bridging between the organic BzA molecules yields a comparably strong binding as the covalent bonds within the octahedral chains. This fundamentally explains the exfoliation potential of BBC despite it featuring only 1D covalent bonds. The required energy for expansions of the unit cell along the three crystallographic axes is calculated using density functional theory (Figure S4c-d). This energy is similar for expansions of up to 10 % along all crystallographic axes. Beyond that, the **a** axis differs from the others. Its expansion energy saturates at about 2.3 eV for expansions of about 20 %, while the other energies continue to increase. This indicates the more solute binding of BBC along the **a** axis and determines the cleavage plane to the **b-c**-plane of the crystal. The calculated crystal structure of the single layer shows only a slight rearrangement of the organic molecules as the unit cell is expanded along the **a** axis. This predicts excellent preservation of the individual layers upon exfoliation. In contrast, the robustness of the 2D layered structure is corroborated by the continuously steep rise in required energies for expansions along the **b** and **c** axes.

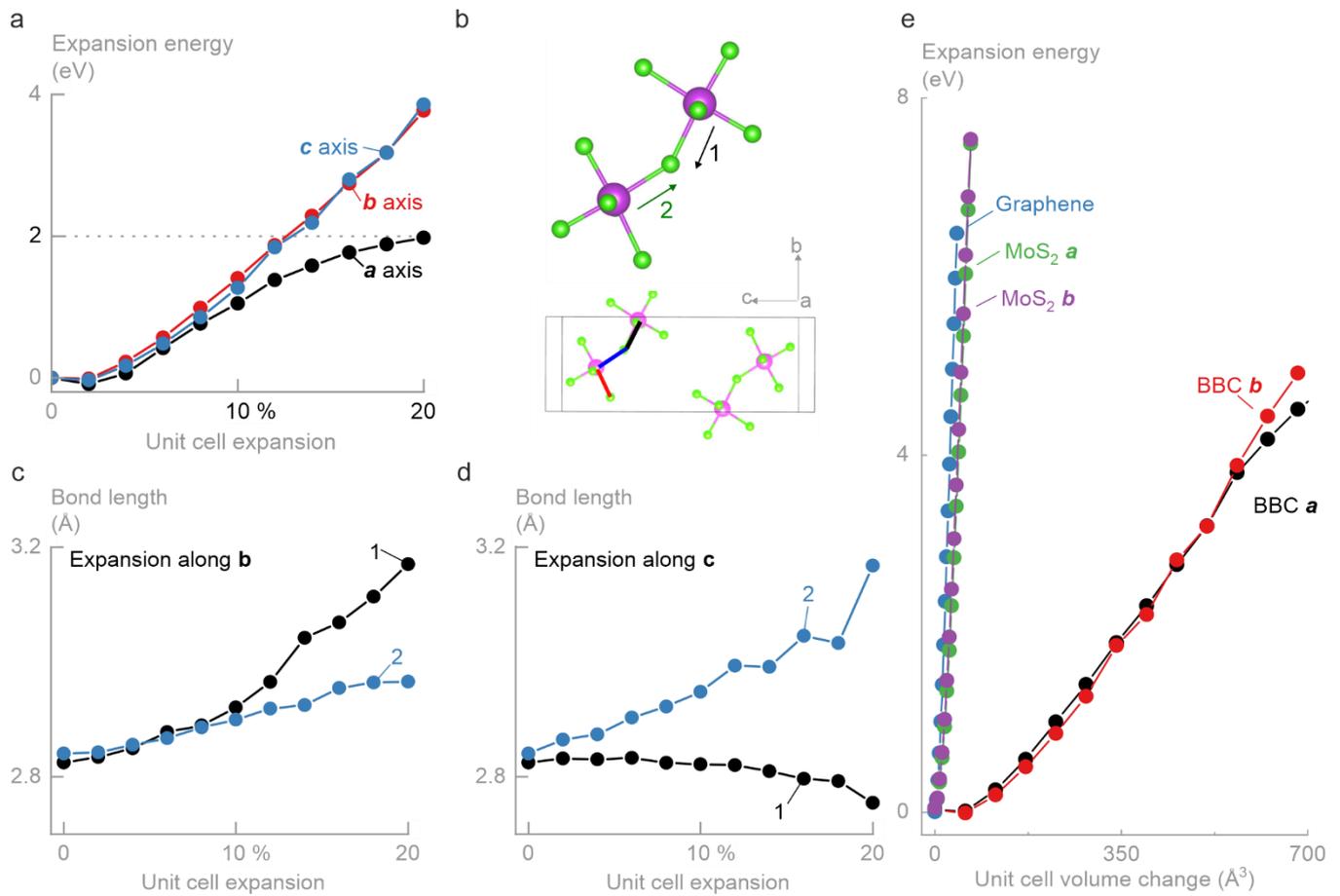

**Figure S4 | Bi-Cl bond length upon expansion of the lattice constants**. **a**, Required energy for expansion of the unit cell along the three crystallographic axes. **b,** Excerpt of the crystal structure with individual Bi-Cl bonds marked. The green (pink) spheres represent Cl (Bi) atoms. **c,** Change of the bond length as a function of the expansion of the unit cell along the ***b*** axis. **d**, Change of the bond length as a function of the expansion of the unit cell along the ***c*** axis. The similarity of results indicates a comparably strong binding along both crystallographic axes ***b*** and ***c***. **e,** Comparison of the 2D sheet binding strength of BBC and other 2D materials.

The required energy for expansion of the unit cells along the crystallographic axes of the 2D sheets of BBC, graphene and MoS$_2$ are shown in Figure S4e. The expansions are performed along the in-plane lattice constants ***b*** and ***c*** for BBC, ***a*** and ***b*** for MoS$_2$, and ***a*** for graphene. Significantly less energy is required to expand the 2D sheet of BBC mirroring the mechanically brittle structure of ultrathin BBC sheets observed in the mechanical exfoliation.

## Density of states and charge density maps

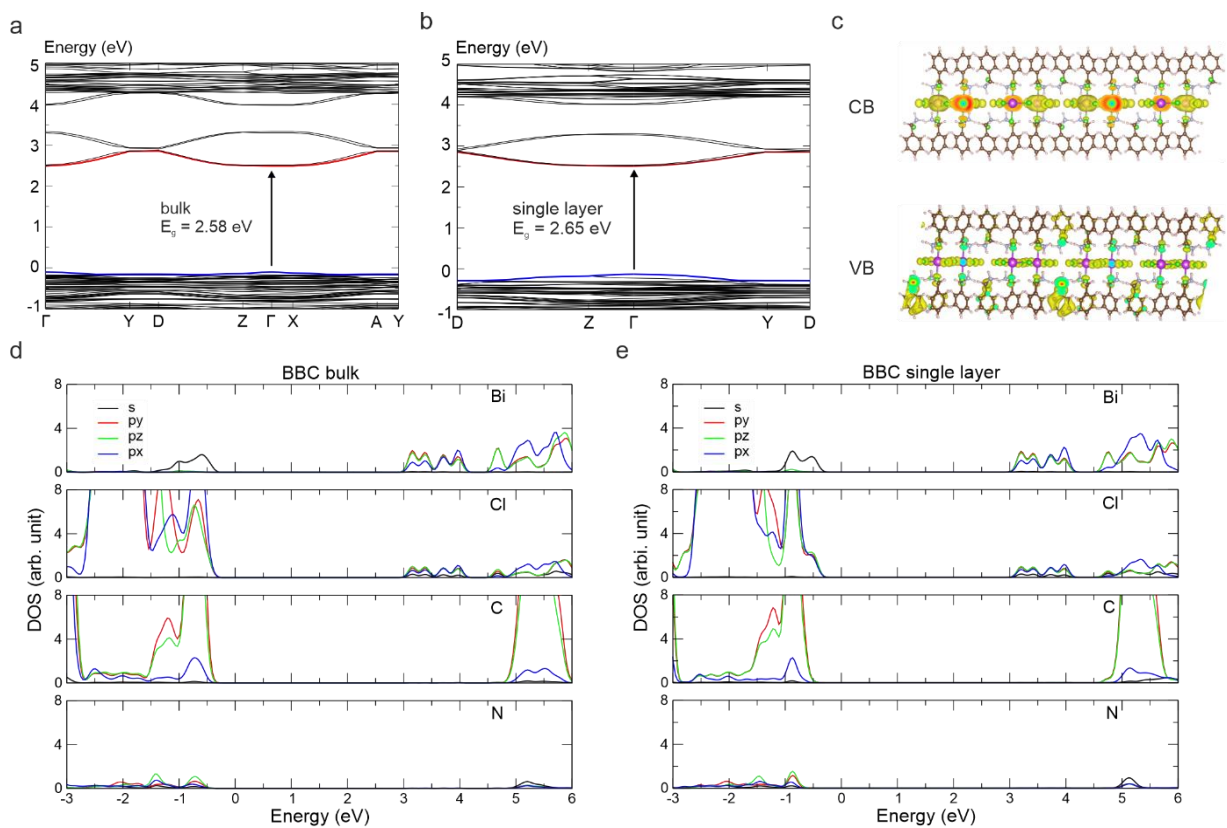

**Figure S5 | Band structure, density of states and charge density maps. a-b**, Electronic band structures of bulk and single layer BBC calculated using the PBE functional. Bulk and single layer BBC exhibit direct band gaps of 2.58 and 2.65 eV at the Γ point. The highest (lowest) valence (conduction) band is highlighted in blue (red). **c**, Conduction band (CB) and valence band (VB) associated with charge density maps of bulk BBC. CB and VB are mainly confined to the inorganic chloro-bismuthate chains. **d-e,** Density of states of bulk and single layer BBC decomposed into orbital and element contributions.

## Stability and long-term storage in air

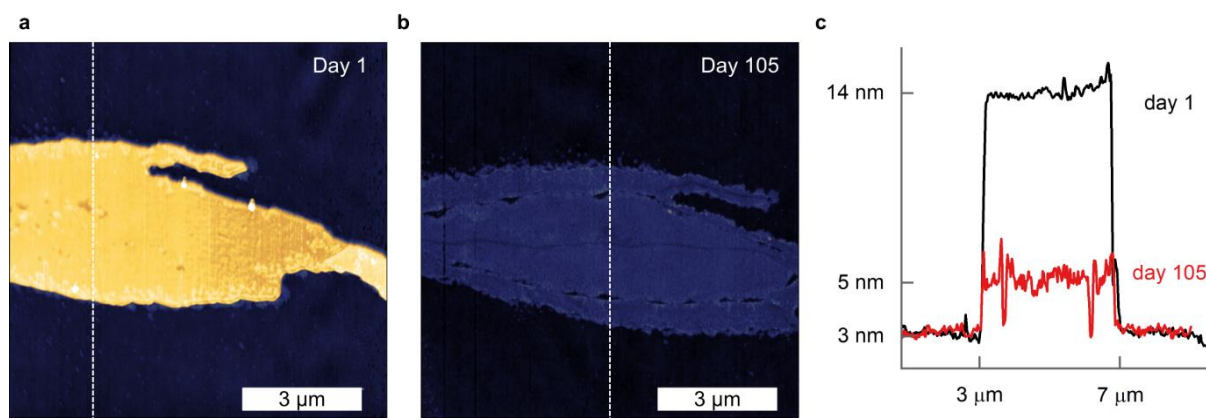

Figure S6 |

**Topological changes of ultrathin BBC under long-term storage in air. a**, AFM image of a typical sample on day of production. **b**, AFM image of the same sample after 105 days of storage in air showing changes in surface topography. **c**, Height profiles of the dashed lines in **a** and **b** showing a thickness decrease from 11 to 2 nm and roughening of the surface.

## Extended optical properties of BBC

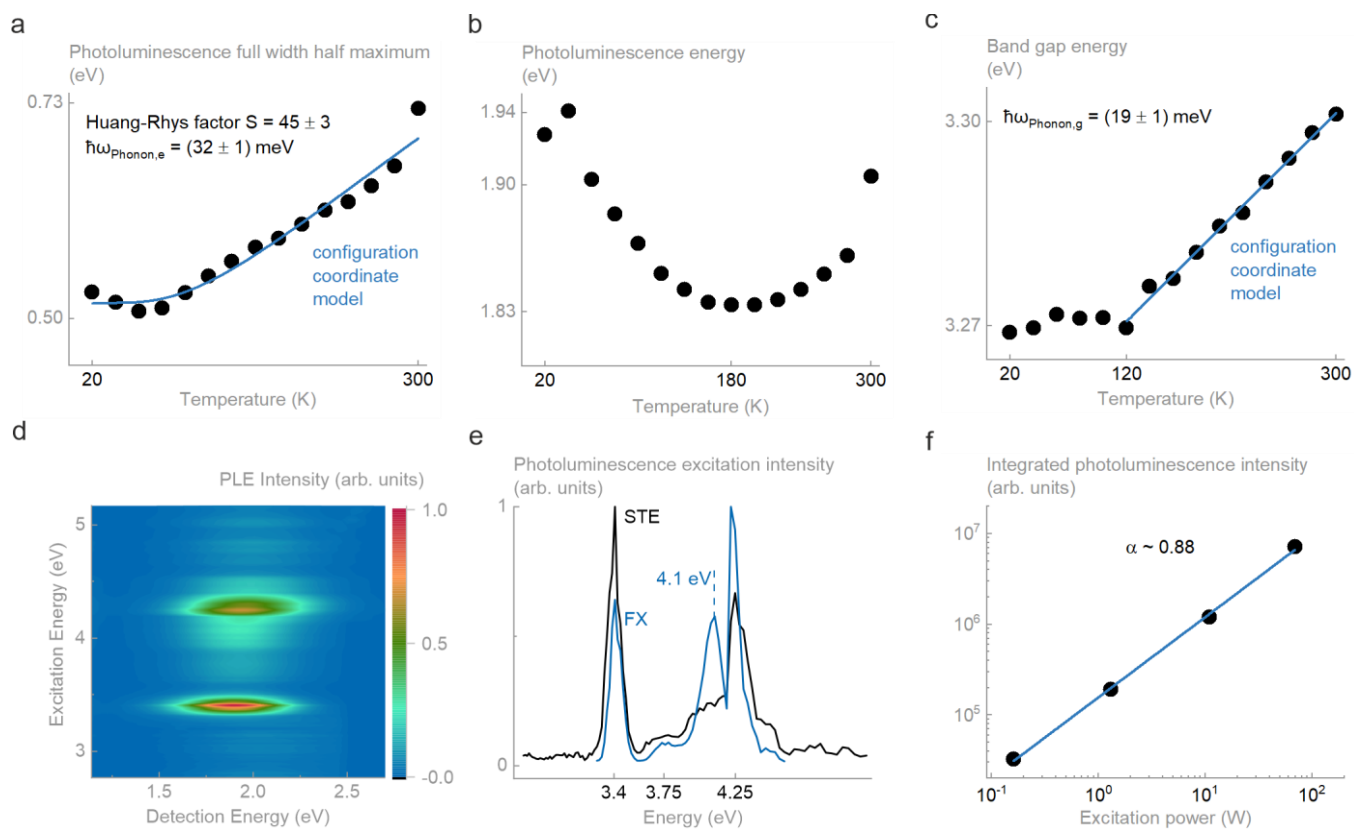

**Figure S7 | Extended optical properties of BBC. a,** Photoluminescence (PL) full width at half maximum (FWHM) as a function of the temperature and fit by the configuration coordinate model (CCM). The FWHM increases following a hyperbolic cotangent yielding a Huang-Rhys factor of S = 45 with the phonon frequency of the excited state $\hbar\Omega_e$ = 32 meV. **b,** PL energy as a function of temperature. It decreases with increasing temperature according to the CCM. The increase after 180 K indicates involvement of another vibrational state. **c,** Onset of absorption as a function of temperate and fit by the CCM. The absorption energy increases linearly from 120 to 300 K yielding the phonon frequency of the ground state $\hbar\Omega_g$ = 19 meV. **d,** 2D plot of PLE spectra of the self-trapped exciton (STE) emission. There is no change in shape or shift of the emission energy for excitation near 3.4 eV. **e,** PLE spectra of the STE (black) and free exciton (FX, blue) emissions. Clearly, both arise from the same states with a strong excitonic resonance at 3.4 eV. Additionally, there is a peak at 4.1 eV in the FX PLE, where no scattering path into the STE exists. The small blue shift of the continuum compared to Figure 3b is due to the higher excitation fluence of 5500 µJ cm$^{-2}$. **f,** Integrated PL intensity as a function of the excitation power. It increases approximately linearly (α = 0.88) indicating that there is no threshold for the onset of PL due to non-radiative recombination through defects at low excitation powers, and no saturation of the absorption at high excitation powers.

**Vibrational properties**

For non-resonant Raman measurements, excitation light from a Nd:YAG laser operating at a wavelength of 532 nm (2.33 eV) was focused into a 200 µm diameter spot with 13 mW excitation power. For resonant Raman measurements, excitation light from a HeCd laser operating at a wavelength of 325 nm (3.82 eV) was focused into a 200 µm diameter spot with 500 mW excitation power. The Raman signal was collected in back-scattering geometry and spectrally filtered through a 1/4 m monochromator (Oriel MS260i) and detected on a Si charge-coupled device camera (Andor iDUS 420). The Si Raman mode at 520 cm$^{-1}$ was used for frequency calibration.

For non-resonant thickness-dependent Raman measurements, a Renishaw inVia Raman microscope with a 785 nm (1.6 eV) laser diode was used for µ-Raman measurements. The beam was focused through a 50X objective achieving a spot size of s ~ 2 µm at 2.16 mW excitation power. The Raman signal was collected by the same objective, dispersed on a diffraction grating (1800 lines cm$^{-1}$, blazed in the visible region) and collected by a CCD camera. The spectral resolution was 1.8 cm$^{-1}$, and the Si Raman mode at 520 cm$^{-1}$ was used for frequency calibration.

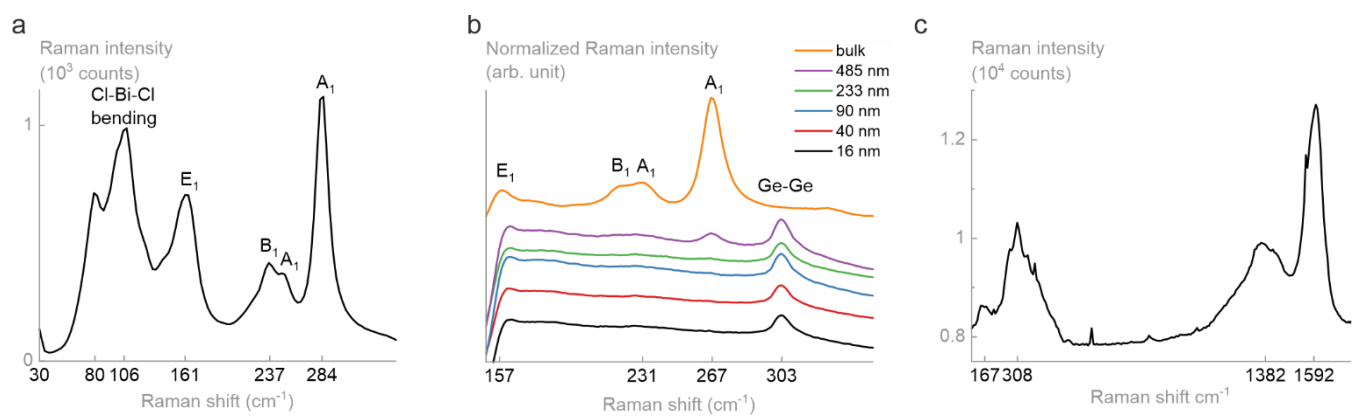

**Figure S8 | Resonant and non-resonant Raman spectroscopy of BBC. a,** Low frequency Raman spectrum of BBC with characteristic Bi-Cl stretching and Cl-Bi-Cl bending vibrations. **b,** Thickness-dependent Raman spectra of BBC. The intensity of the peaks decreases with the sample thickness and eventually vanishes for ultrathin sheets. **c,** Resonant Raman spectrum of BBC showing significant differences to the non-resonant excitation.

The Raman spectra are typical for 1D structures with an octahedral arrangement of halogen atoms with two chlorine bridges linking two adjacent bismuth atoms. We attribute the peaks at 284, 249, 237 and 161 cm$^{-1}$ to Bi-Cl stretching vibrations (2A$_1$, B$_1$, E$_1$) and the peaks at 106 and 80 cm$^{-1}$ to Ci-Bi-Cl bending vibrations[27]. The peaks shift to lower Raman frequencies when a 785 nm laser was used for excitation. The intensity of the peaks decreases with the sample thickness and eventually vanishes for ultrathin sheets indicating weak Raman scattering efficiencies. We assign the peak at 303 cm$^{-1}$ to the second order acoustic phonon in silicon, which is supported by the evolution of the peak as it disappears for very thick samples.

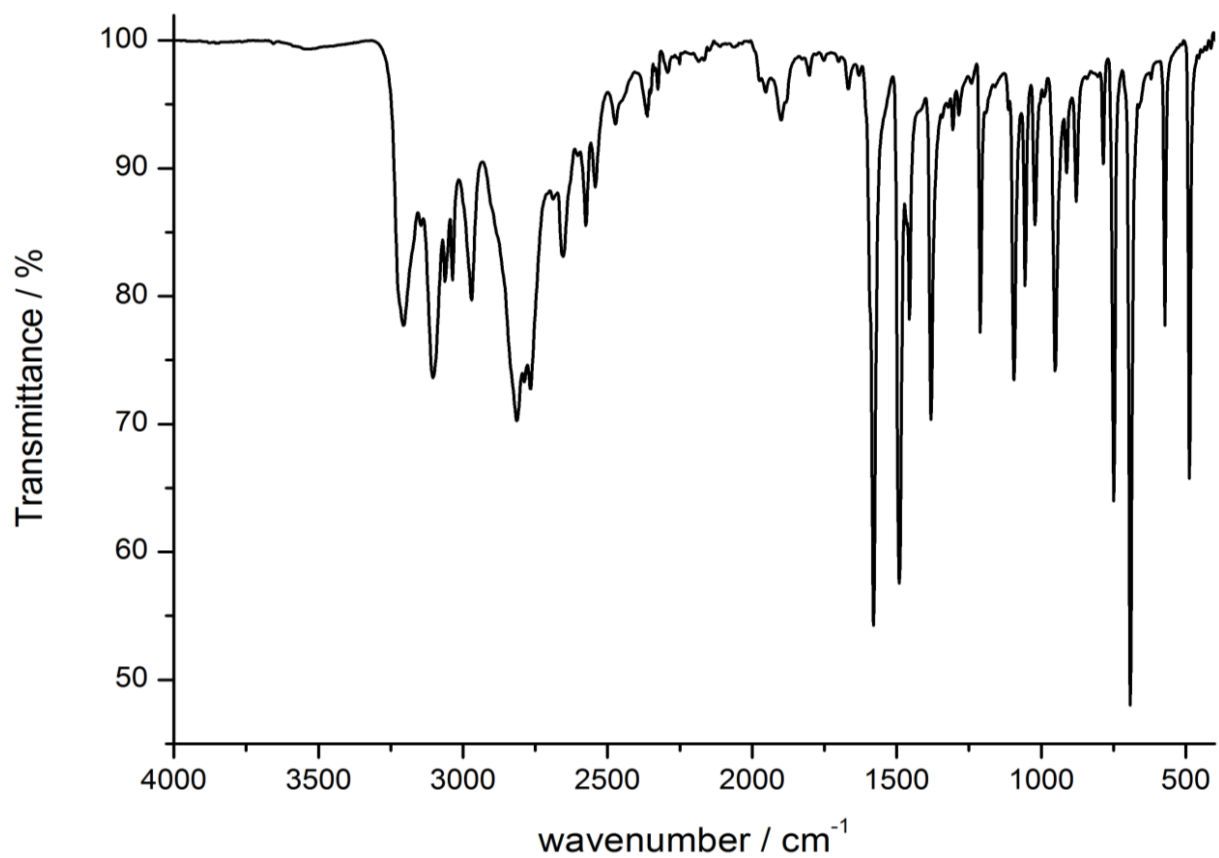

**Figure S9 | IR spectrum of BBC.** An IR spectrum of BBC was recorded on a Bruker Tensor 37 FT-IR spectrometer equipped with an ATR-Platinum measuring unit. The typical pattern of the benzylammonium moiety is observed.[12]

**References**

[1] Kresse, G. & Joubert, D. From ultrasoft pseudopotentials to the projector augmented-wave method. *Phys. Rev. B* **59**, 1758 (1999).

[2] Kresse, G. & Hafner, J. Ab initio molecular dynamics for liquid metals. *Phys. Rev. B* **47**, 558 (1993).

[3] Kresse, G. & Hafner, J. Ab initio molecular-dynamics simulation of the liquid-metal–amorphous-semiconductor transition in germanium. *Phys. Rev. B* **49**, 14251 (1994).

[4] Perdew, J., Burke, K. & Ernzerhof, M. Generalized gradient approximation made simple. *Phys. Rev. Lett.* **77**, 3865 (1996).

[5] Tkatchenko, A. & Scheffler, M. Accurate molecular van der Waals interactions from ground-state electron density and free-atom reference data. *Phys. Rev. Lett.* **102**, 073005 (2009).

[6] Bruker, APEX3. Bruker AXS Inc., Madison, Wisconsin, USA (2018).

[7] a) Sheldrick, G. M. A short history of SHELX. *Acta Cryst. A* **64**, 112-122 (2008); b) Sheldrick, G. M. SHELXT – Integrated Space-Group and Crystal-Structure Determination. *Acta Cryst. A* **71**, 3-8 (2015); c) Sheldrick, G. M. Crystal Structure Refinement with SHELXL. *Acta Cryst. C* **71**, 3-8 (2015).

[8] Dolomanov, O. V., Bourhis, L. J., Gildea, R. J., Howard, J. A. K., Puschmann, H. OLEX2: A Complete Structure Solution, Refinement and Analysis Program. *J. Appl. Crystallogr*. **42**, 339-341 (2009).

[9] Brandenburg, K. Diamond, Crystal Impact GbR: Bonn, Germany (2005).

[10] Mousdis, G. A., Papavassiliou, G. C., Terzis, A. & Raptopoulou, C. P. Preparation, Structures and Optical Properties of [$H_3N(CH_2)_6NH_3$]Bi$X_5$(X=I, Cl) and [$H_3N(CH_2)_6NH_3$]Sb$X_5$ (X=I, Br). *Z. Naturforsch.* **53b**, 927-931 (1998).

[11] Macrae, C. F., Edgington, P. R., McCabe, P., Pidcock, E., Shields, G. P., Taylor, R., Towler, M. & van de Streek, J. Mercury CSD 2.0 - new features for the visualization and investigation of crystal structures. *J. Appl. Cryst.* **39**, 453-457 (2006).

[12] Brittain, H. G. Vibrational Spectroscopic Studies of Cocrystals and Salts. 4. Cocrystal Products formed by Benzylamine, α-Methylbenzylamine, and their Chloride Salts. Crystal G. & Des. **116**, 2500-2509, (2011).